\documentclass[12pt]{iopart}

\usepackage{graphicx}
\usepackage{ccaption}
\begin{document}

\title[PHENIX Studies of the Scaling Properties of Elliptic Flow at RHIC]{PHENIX Studies of the Scaling Properties of Elliptic Flow at
   RHIC}

\author{A. Taranenko for the PHENIX Collaboration\footnote{For the full list of PHENIX authors and acknowledgements, see Appendix
'Collaborations' of this volume.}
}

\address{Department of Chemistry, SUNY , Stony Brook NY 11794}
\ead{arkadij@rcf.rhic.bnl.gov}
\begin{abstract}
	Recent PHENIX elliptic flow ($v_2$) measurements for identified particles produced 
in Au+Au and Cu+Cu collisions at $\sqrt{s_{NN}}=200$~GeV are presented and compared 
to other RHIC measurements. They indicate universal scaling of $v_2$ compatible with 
partonic collectivity leading to the flow of light, strange and heavy quarks with a 
common expansion velocity field.
\end{abstract}


\section{Introduction}
Universal scaling of elliptic flow ($v_2$) has been recently observed at 
RHIC  \cite{v2scaling,v2scaling1,v2scaling2}. That is, for a broad range 
of  particle species, $v_2/ n_q$ vs. $KE_T/n_q$ scales to a single function;
here, $n_q$ and $KE_T$ are  
the number of valence quarks ($n_q =2, 3$ for mesons and baryons respectively) and 
the transverse kinetic energy of the particle. This observation has been
interpreted as evidence that transverse expansion of the matter produced in 
energetic RHIC collisions, occurs during a phase dominated by partonic 
collectivity. 

     In this contribution we further demonstrate; 
(i) the partonic origin of $v_2$ via scaling tests 
for the $\phi$ meson; (ii) essentially full thermalization 
for the charm quark via scaling tests for D meson $v_2$; 
and (iii) validate universal scaling of $v_2$ in Au+Au and Cu+Cu 
collisions at $\sqrt{s_{NN}}=200$~GeV.

\section{Further tests for universal scaling of elliptic flow at RHIC}
Differential $v_2$ measurements for multi-strange hadrons (~$\phi$, $\Xi$, $\Omega$~)
allow us to address the important question of how the existence of a hadronic
phase affects $v_2$. These particles are  to have a small hadronic cross section 
for interaction with non-strange hadrons. 
Consequently, if elliptic flow was established in a phase involving hadrons interacting 
with their standard hadronic cross sections (post-hadronization), one would 
expect $v_2$ for the multi-strange hadrons  to be significantly smaller than that for  
other hadrons (e.g. $p$, K, $\pi$). The left panel of Fig.~1 compares PHENIX $v_2$ results 
for the $\phi$ meson, $\pi^{\pm}$, $K^{\pm}$ and $(\overline{p})p$ from 20-60\% 
central Au+Au collisions; they show that, despite its mass which 
is comparable to that for the $p$ and $\Lambda$, its $v_2(KE_T)$ values are  
similar to those for the lighter mesons ($\pi$ and $K$)  
whose hadronic re-scattering cross sections are expected to be large. 
The right panel of Fig.~1  shows the positive results from a validation 
test for universal scaling of the $\phi$ meson $v_2$. The recent high-statistics 
preliminary STAR results for $\Omega$ and $\Xi$ produced in Au+Au collisions at 
$\sqrt{s_{NN}} = 200$~GeV \cite{v2star2} show a similar universal scaling \cite{v2scaling2}.
These results suggest that the $v_2$ values for hadrons comprised of u, d and s quarks
develop in the partonic phase and the existence of a hadronic phase does  not lead 
to their modification.

\begin{figure}[tbh]
\begin{center}
\includegraphics[width=0.84\textwidth,height=0.24\textheight,clip=]{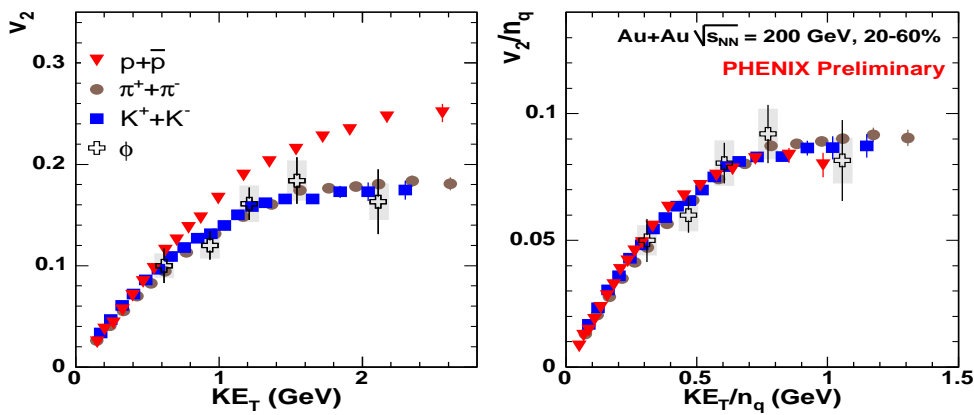}
\caption{$v_2$ vs. $KE_T$. and $v_2/n_q$ vs. $KE_T/n_q$ for $\pi, K, p$ and $\phi$ 
mesons detected in semi-central (20-60\%) Au+Au collisions at $\sqrt{s_{NN}} = 200$~GeV. }
\end{center}
\end{figure}

  Further insight into properties of the produced medium at RHIC
can be obtained  from the study of elliptic flow of 
particles comprised of heavy quarks (charm or bottom).
If charm quarks flow like light quarks,
it would indicate an unexpectedly strong interaction of charm with the medium. 
One method of accessing charm $v_2$ is to measure the $v_2$ for  
``non-photonic" electrons \cite{v2D}. A recent high statistics PHENIX measurements show rather
larger elliptic flow for ``non-photonic" electrons from Au+Au collsions at $\sqrt{s_{NN}}=200$ GeV \cite{v2D}.  
The ``non-photonic" electrons  arise mainly from D meson decay for $p_T<$ 2GeV/c. Therefore, the large heavy
flavor electron $v_2$ indicates that the D meson and consequently the 
charm quark, do flow with a large $v_2$. This suggests that the in-medium 
interactions are strong and perhaps frequent enough to thermalize the 
charm quark. If this is indeed the case, then $v_2(KE_T)$ for the D meson should 
show the same universal scaling presented earlier for other hadrons.

The  $v_2$ values for D mesons can been obtained from the values for non-photonic electrons 
via detailed simulations, see \cite{v2D} for details.
For these simulations, it is necessary to assume the shape for $v_2(p_T)$ for the D mesons.
The left and right panels of Fig.~2  compare the unscaled and scaled 
results (respectively) for $v_2$ vs. $KE_T$ for $\pi, K, p$ and $D$ mesons measured 
in minimum bias  Au+Au collisions at $\sqrt{s_{NN}}=200$ GeV. 
Here, the assumed shape for the $p_T$ dependence of the D meson $v_2$ is that for 
the proton (see \cite{v2scaling2,v2D}).
\begin{figure*}[tb]
\begin{center}
\includegraphics[width=0.9\linewidth, height=0.24\textheight,clip]{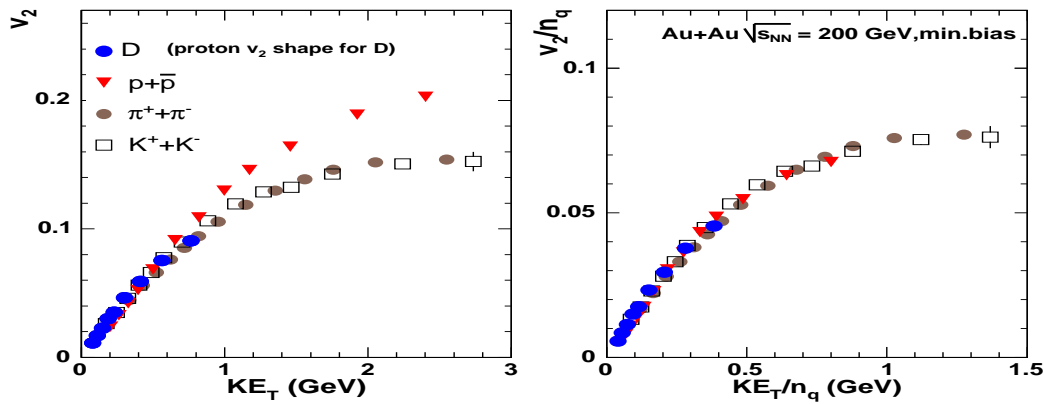} 
\caption{$v_2$ vs. $KE_T$. and $v_2/n_q$ vs. $KE_T/n_q$. 
for $\pi^{\pm}, K^{\pm}, (\bar{p})p$ from  minimum bias Au+Au 
collisions at $\sqrt{s_{NN}}=200$ GeV \cite{v2scaling}. 
Results for D-meson is shown for the proton-like  assumption for the shape of  D $v_2(p_T)$ \cite{v2scaling2,v2D}}
\label{fig:v2_Raa}
\end{center}
\end{figure*} 
This assumption is compatible with a large array of differential
$v_2$  measurements for different ``heavy" particle species \cite{v2scaling,v2scaling2,v2star}. 
Both panels of Fig.~2  show robust scaling; they confirm 
that the D mesons (for $p_T < 2.0$ GeV/c) and hence, their associated charm quarks, flow with 
a fluid velocity field similar to that for the other quarks. 
This suggests that the medium responds as a thermalized fluid and the 
transport mean free path is small.

	If $v_2$ develops in a phase dominated by partonic collectivity and $u$, $d$, $s$ and 
$c$ quarks flow with a common velocity field, one expects universal scaling for all particle 
species measured at RHIC. 
Figure 3(a) shows differential flow measurements $v_2(p_T)$, for several particle species
produced at mid-rapidity in central and semi-central Au+Au collisions at  $\sqrt{s_{NN}}$=200 GeV; they
span essentially the full range of measurements (several hundred data points) at RHIC. The $v_2$
values  for pions ($\pi^{\pm},\pi^{0}$), kaons $K^{\pm}$, (anti-)protons ($\bar{p}$,p),
(anti-)deuterons ($\bar{d}$)d and the $\phi$ meson comprise results from the 
PHENIX \cite{v2scaling,v2scaling1,v2scaling2,v2phenix2}.  The values for neutral kaons ($K^{0}$),
lambdas ($\Lambda+\bar{\Lambda}$), cascades ($\Xi+\bar{\Xi}$) and omegas ($\Omega+\bar{\Omega}$)
represent  results from the STAR \cite{v2star2,v2star}. 
Fig. ~3~(b) shows the scaled results ($v_2/n_q \epsilon$ vs $KE_T/n_q$)  obtained from 
the same data; here $\epsilon$ is the integral $v_2$ of charged hadrons 
for each of the indicated centrality selections, multiplied by a constant factor $k\sim 3.2$
(i.e. $\epsilon = k\times v_2$) \cite{v2scaling,v2scaling1,v2scaling2}. 
Recent measurements \cite{v2scaling} indicate that $v_2(p_T)/\epsilon$ is independent of 
centrality and the size of the colliding system as would be expected from a 
hydrodynamic system.  Fig. 3(b) indicates that the relatively complicated 
dependence of $v_2$ on centrality, transverse momentum, 
particle type, etc., for particles produced at mid-rapidity  can be scaled 
to a single function. Fig. 1(c) demonstrates that the same scaling also holds 
for $\pi^{\pm}$,($K^{\pm}$) and ($\bar{p}$,p) produced at 
mid-rapidity in Cu+Cu collisions at $\sqrt{s_{NN}}$=200 GeV.

\begin{figure}[tbh]
  \begin{minipage}{0.46\textwidth}
    \includegraphics*[width=0.85\textwidth]{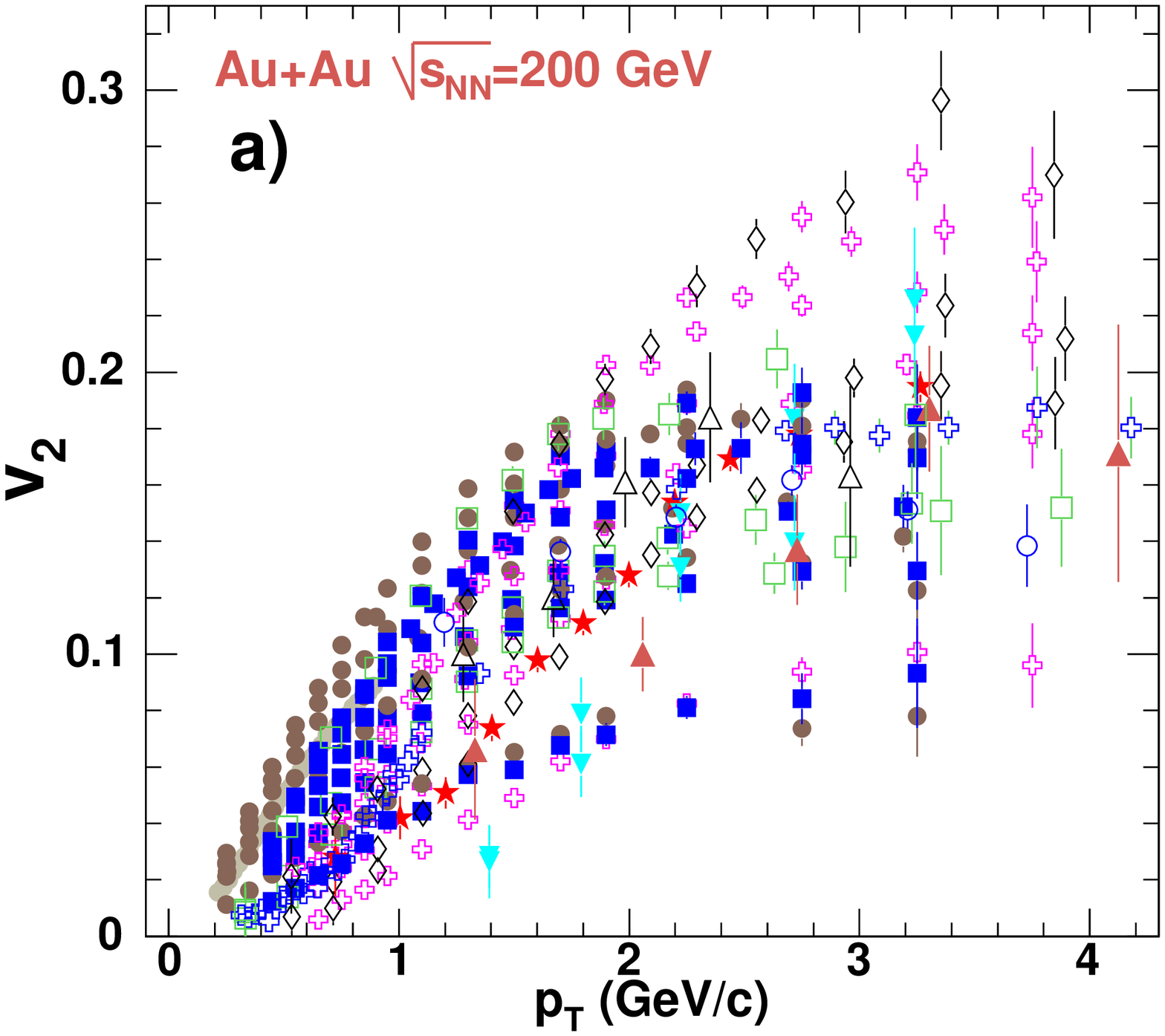}
  \end{minipage}
  \begin{minipage}{0.46\textwidth}
 \includegraphics*[width=0.85\textwidth]{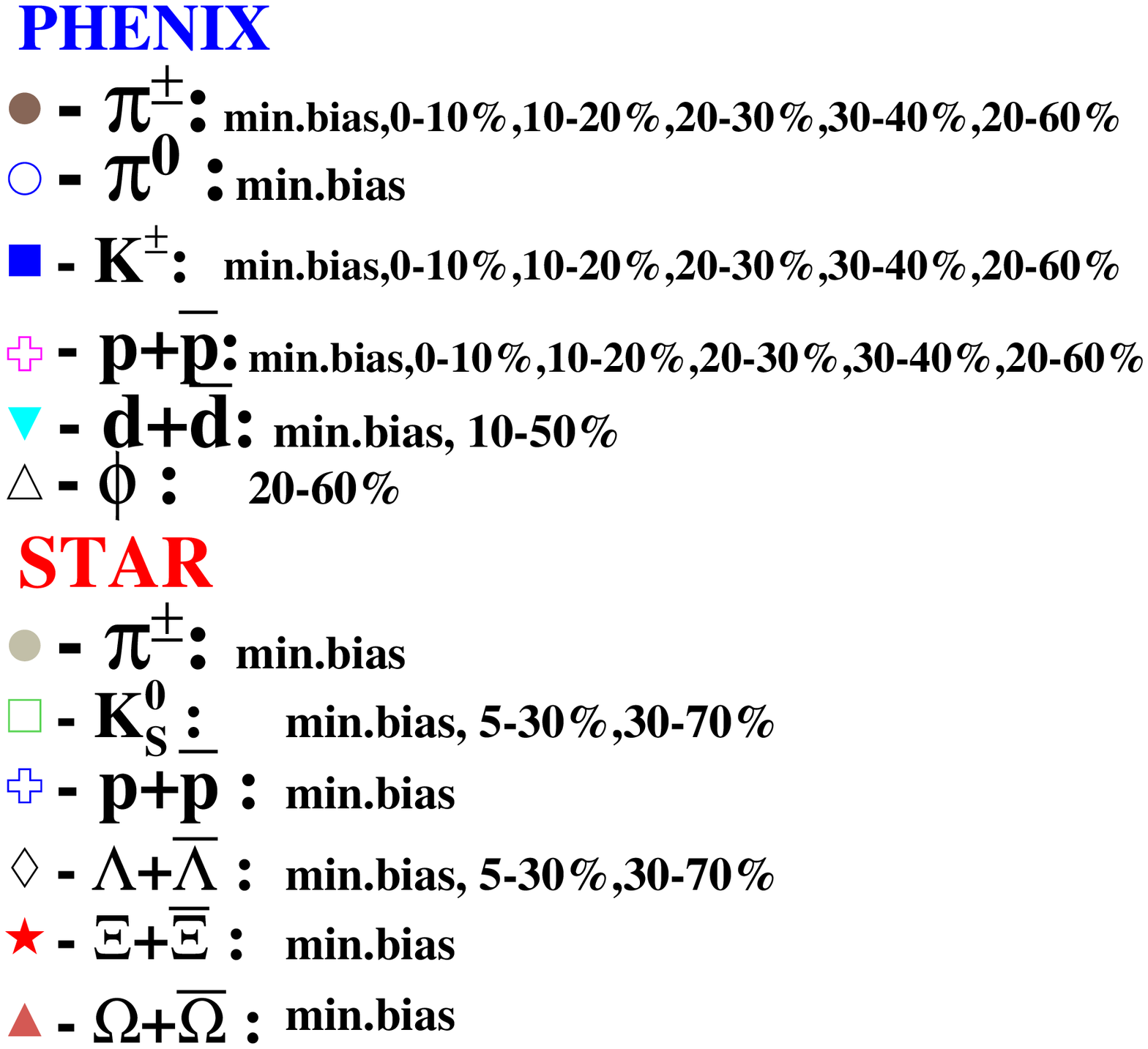}
\end{minipage}
  \begin{minipage}{0.46\textwidth}
    \includegraphics*[width=0.85\textwidth]{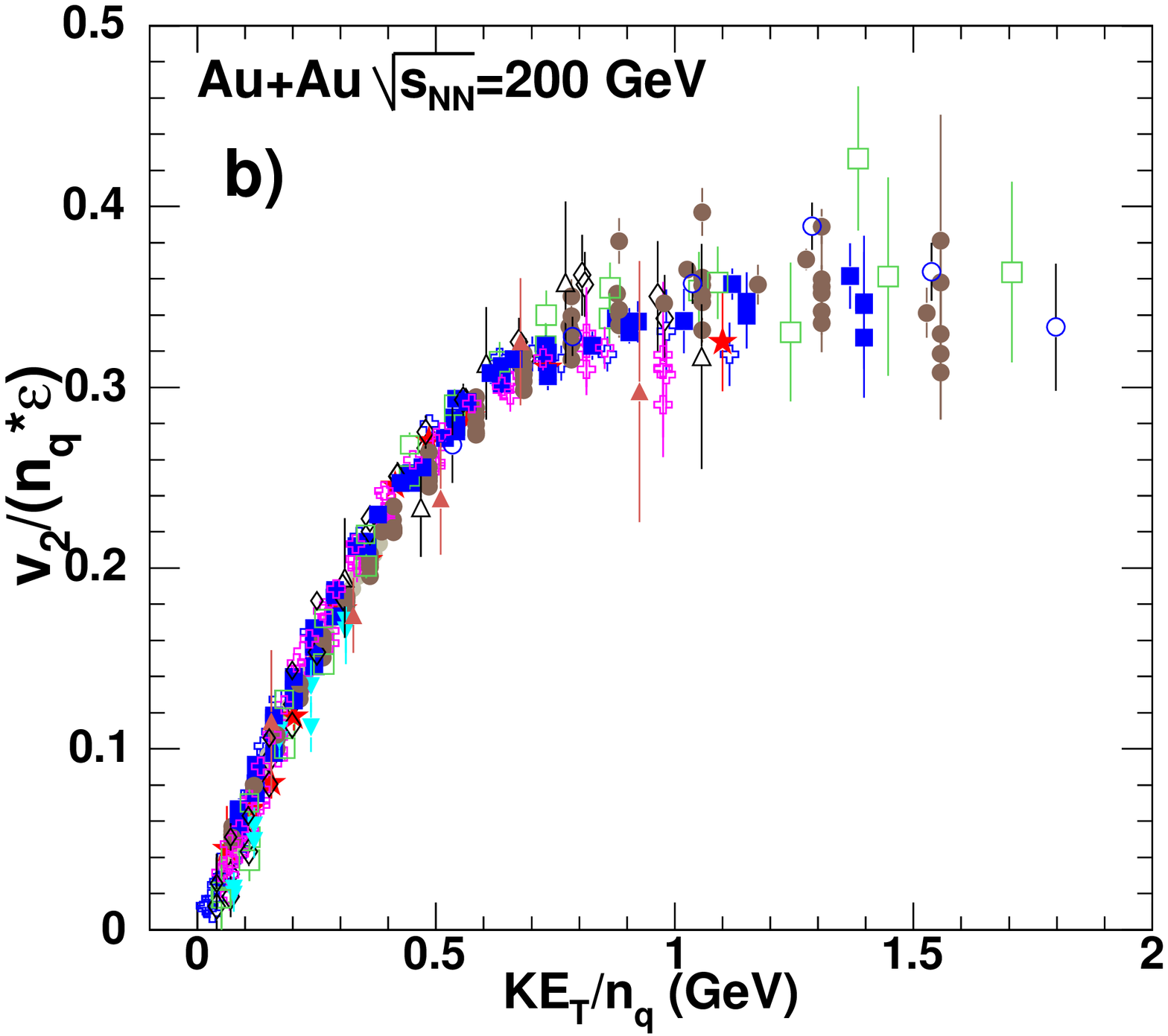}
  \end{minipage}\hfill
\begin{minipage}{0.46\textwidth}
    \includegraphics*[width=0.85\textwidth]{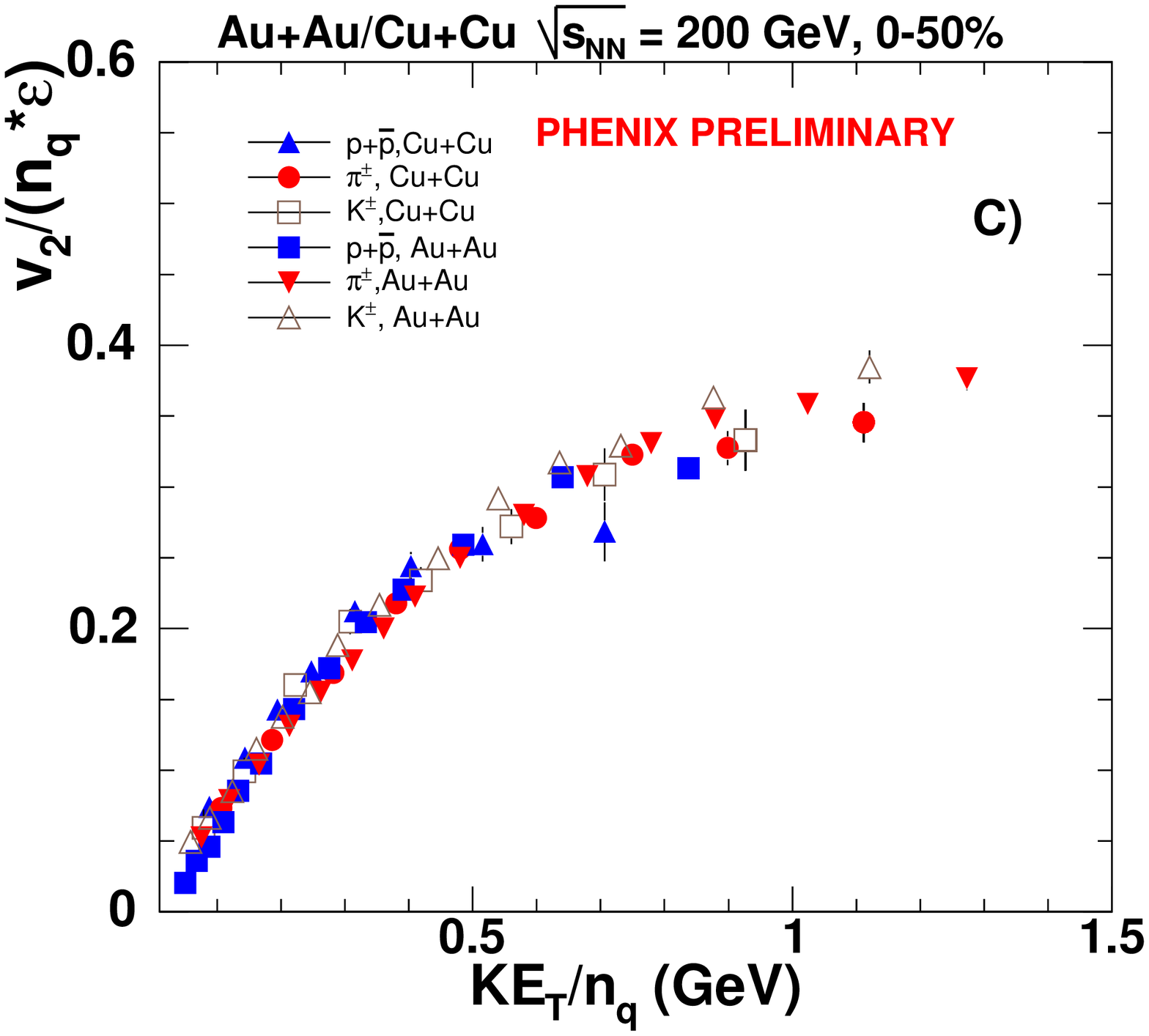}
\end{minipage}
\caption{ (a) $v_2$ vs. $p_T$  and (b) $v_2/(n_q \epsilon$) vs. $KE_T/n_q$ 
for several centralities and particle species 
as indicated.
(c) $v_2/(n_q \epsilon$) vs. $KE_T/n_q$ for 
$\pi^{\pm}$, $K^{\pm}$, and ($\bar{p}$,p) from Cu+Cu and  Au+Au collisions at $\sqrt{s_{NN}} = 200$~GeV }
\label{fig:dphi_dists}
\end{figure}

\section{Summary}

 Validation tests for the universal scaling of $v_2$ at RHIC, suggest that 
the transverse expansion dynamics leading to elliptic flow are not controlled by 
ordinary hadrons interacting with their standard hadronic 
cross sections. Instead, they suggest  a pre-hardonization state exhibiting 
partonic collectivity leading to the flow of light, strange and heavy quarks 
with a common expansion velocity field.

\end{document}